\documentclass{ifacconf}

\usepackage{amsmath,amssymb,amsfonts}
\usepackage{algorithm}
\usepackage{algpseudocode}
\usepackage{natbib}        
\usepackage{graphicx}      
\usepackage{xcolor}
\usepackage{multirow}

\newtheorem{theorem}{Theorem}

\newtheorem{lemma}[theorem]{Lemma}

\newtheorem{definition}[theorem]{Definition}

\newenvironment{proof}{\textit{Proof:}}{\hfill$\square$}

\newcommand{\End}{\hfill $\square$}

\usepackage{subfiles} 

\begin{document}
\begin{frontmatter}

\title{Data-driven MPC of descriptor systems: \\A case study for power networks}


\author[First]{Philipp Schmitz} 
\author[Second]{Alexander Engelmann} 
\author[Second]{Timm Faulwasser}
\author[First]{Karl Worthmann}

\address[First]{Institute of Mathematics, Optimization-based Control, Technische Universität Ilmenau, Ilmenau, Germany\\
	 \{philipp.schmitz,karl.worthmann\}@tu-ilmenau.de.}
\address[Second]{Institute for Energy Systems, Energy Efficiency and Energy Economics, TU Dortmund University, Dortmund, Germany \\ \{alexander.engelmann, timm.faulwasser\}@ieee.org.}

\begin{abstract}                
Recently, data-driven predictive control of linear systems has received wide-spread research attention. It hinges on the fundamental lemma by Willems et al. In a previous paper, we have shown how this framework can be applied to predictive control of linear time-invariant descriptor systems. 
In the present paper, we present a case study wherein we apply data-driven predictive control to a discrete-time descriptor model obtained by discretization of the power-swing equations for a nine-bus system. Our results shows the efficacy of the proposed control scheme and they underpin the prospect of the data-driven framework for control of descriptor systems.  
\end{abstract}


\end{frontmatter}

\section{Introduction}

Rotor-angle and frequency stability are crucial for a safe  power system operation.
They describe the ability of the power system to preserve synchrony of  generators and to bound frequency   oscillations after load changes or failures of subsystems \citep{Kundur2004}.

A classical approach for achieving these goals is  hierarchical Automatic Generation Control (AGC).
AGC typically consist of a fast inner droop control, where proportional controllers counteract the disturbance and an outer secondary control loop, which ensures asymptotic convergence to the nominal frequency as well as load sharing among generators \citep{Ibraheem2005,Li2016}. 
However, AGC schemes need a model of the generator dynamics and the transmission lines for controller design. 
In practice, model parameters are often unknown or may change over time, e.g., due to temperature drifts. 
Moreover, AGC schemes often neglect input constraints, which might lead to difficulties especially in case of large disturbances in  small grids with limited power generation.

Model Predictive Control (MPC) has the potential to include technical limits such as voltage bounds  \citep{Liu2019}.
Moreover, \emph{data-driven} MPC schemes are able to use measured data  directly  instead of a system model, which avoids a potentially costly system identification step.
Furthermore, they are able to  adapt to slow parameter drifts \citep{Berberich2020b,Bilgic2022}.

Data-driven MPC schemes for frequency  control have  been proposed by \cite{Wang2019a,Huang2021}.
However, for these schemes rigorous stability properties have---to the best of our knowledge---so far not been established either due to nonlinearities in the system or due to the fact that the system is described by a descriptor system rather than a discrete- or continuous-time linear system, e.g., governed by an ordinary differential equations.

In the present paper, we use rely on results from our previous work \citep{Schmitz2022}, wherein we tailor Willem's fundamental lemma to Linear Descriptor Systems~(LDSs).
Therein, we  established closed-loop stability guarantees for LDSs  controlled by a data-driven MPC scheme. 
In the present work, we apply these results for frequency control in power systems.
Specifically, we illustrate the performance of data-driven predictive control on a 9-bus system with three generators.  

This work is organized as follows: Section~2 describes the dynamic power system model as a descriptor system with the underlying assumptions. Section~3 continues with the fundamental lemma for descriptor systems, followed by Section~4 focusing on data-driven predictive control. In Section~5 this data-driven control scheme is applied to frequency control for a 9-bus system.

\noindent \textbf{Notation}: $\mathbb{N}_0$, $\mathbb{N}$ denote the natural numbers with and without zero, respectively. Moreover, for two numbers $a,b \in \mathbb{N}_0$ with $a \leq b$, the non-empty interval $[a,b] \cap \mathbb{N}_0$ is denoted by $[a:b]$. %

For a matrix $A$ the entry of the $i$th column and the $j$th row is denoted by $A_{i,j}$. The identity $\mathbb R^{n\times n}$ is denoted by $I_n$.

For a function $u:\Omega \to \Gamma$, we denote the restriction of $u$ to $\Omega_0 \subset\Omega$ by $u|_{\Omega_0}$. 
Considering a map $u : [t:T-1] \rightarrow \mathbb R^{k}$ with $t< T$, we denote the vectorization of $u$ by 
\begin{equation*}
    \mathbf{u}_{[t,T-1]} \doteq \begin{bmatrix}
    f(t)^\top & 
    \hdots & f(T-1)^\top
    \end{bmatrix}^\top\in \mathbb R^{k(T-t)}
\end{equation*}
and, for $L \in \mathbb{N}$ with 
$L \leq T-t$, the corresponding Hankel matrix~$H_{L} (\mathbf{u}_{[t,T-1]}) \in\mathbb R^{kL \times (T-t-L+1)}$ is defined by
\begin{equation*}
   H_{L} (\mathbf{u}_{[t,T-1]}) \doteq  \begin{bmatrix}
        f(u) 
        & \dots & u(T-L)\\
        \vdots & \ddots & \vdots\\
        u(t+L-1) 
        & \dots & u(T-1)
    \end{bmatrix}.
\end{equation*}
Given a symmetric positive-definite matrix $Q$ we define the norm $\lVert x\rVert_{Q}\doteq (x^\top Q x)^{1/2}$.

\section{Problem Formulation}
	Consider a power system with $\mathcal N =\{1,\dots,n\}$ buses and $\mathcal G \subseteq \mathcal N$ generators. Without loss of generality we assume $\mathcal G = \{1,\dots,g\}$.
	The grid parameters are described by the bus admittance matrix $Y=\bar G+\mathrm{i}\bar B \in \mathbb{C}^{n\times n}$, where $\bar G \in \mathbb R^{n\times n}$ describes the conductances and $\bar B \in \mathbb R^{n \times n}$ describes the susceptances of all transmission lines.
	We make the following assumptions:
	\begin{enumerate}
	    \item [a)] lossless system, i.e., $\bar G=0$
	    \item [b)] second-order synchronous generator model
	    \item [c1)] constant power loads for all nodes
	    \item [c2)] constant bus voltages
	    \item [d)] small voltage angle differences
	    \item [e)] connected system topology
	\end{enumerate}
	These assumptions are quite specific but also common in the context of frequency control, cf. \citep{Simpson-Porco2013,Schiffer2016,Song2015}.
	
	Following \cite{Bergen1981}, the dynamics of a synchronous generator $i \in \mathcal{G}$ are described by 
	\begin{align}\label{eq:genDyn}
		M_i \ddot \theta_i + D_i \dot \theta_i = p_i - p_i^d - \sum_{j \in \mathcal{N}} \bar B_{i,j} \sin(\theta_i - \theta_j).
		\end{align}
	Here, $M_i>0$ is the inertia of the generator, $D_i>0$ is the damping, $p_i\geq 0$ is the mechanical power provided to the generator, $p_i^d\geq 0$ is the active power demand, and $\theta_i\in\mathbb R$ is the phase angle. 
	We rewrite \eqref{eq:genDyn} as 
	first-order system
	\begin{equation*}\label{eq:genModel}
		\begin{bmatrix}
			\dot \theta_i \\
			\dot  \omega_i 
		\end{bmatrix}
		\hspace{-1mm}= \hspace{-1mm}
		\begin{bmatrix}
			\omega_i \\
			\hspace{-.5mm}M_i^{-1} \Big ( p_i - p_i^d \hspace{-.7mm} - \hspace{-.7mm}\sum_{j \in \mathcal{N}} \bar B_{i,j} \sin(\theta_i \hspace{-.5mm} - \hspace{-.3mm} \theta_j) -D_i  \omega_i \Big ) \hspace{-.7mm}
		\end{bmatrix} \hspace{-1mm}.
	\end{equation*}
	For all nodes without a generator, we assume constant power loads, cf. assumption c1).
	This leads to the algebraic equations
	\begin{align*}
	 - p_i^d - \sum_{j \in \mathcal{N}} \bar B_{i,j} \sin(\theta_i - \theta_j) =0  \qquad \text{for all}\ i \in \mathcal{N} \setminus \mathcal G.
	\end{align*}

\textcolor{blue}{}Next, we derive an approximation of this nonlinear system invoking assumption d). To this end, we first linearize $\sin(\theta_i-\theta_j) \approx \theta_i-\theta_j$ \citep{Liu2021,Zhao2014}. This leads, together with Euler-forward discretization, to the discrete-time LDS
\begin{subequations} \label{sys}
	\begin{align}
		\label{sysa}
		\tag{\theequation{}a}
		Ex(t+1) &= Ax(t) + Bu(t) + Fw(t), \\
		\label{sysb}
		\tag{\theequation{}b}
		y(t) &= Cx(t)
	\end{align}
\end{subequations}
with matrices $E$,~$A\in\mathbb R^{(n+g)\times (n+g)}$, $B\in \mathbb R^{(n+g)\times g}$ and $C\in\mathbb R^{m\times(n+g)}$.
Let $L$ be the Laplacian of the admittance matrix $\bar B$, i.e.
\begin{equation*}
    L = \operatorname{diag}\left(\sum_{j\neq 1} \bar  B_{1,j},\dots, \sum_{j\neq n} \bar  B_{n,j}\right) - \bar  B.
\end{equation*}
Then
\begin{align*}
E &= \begin{bmatrix}
		0 & I_g & 0\\
		I_g & 0 & 0\\
		0 & 0 & 0
	\end{bmatrix},&
 A &= E - \tau \begin{bmatrix}
    I_g & 0 & 0\\
    0 & \tilde M & 0\\
    0 & 0 & I_{n-g}
\end{bmatrix}\left[
\begin{array}{ccc}
    I_{g} & 0 & 0\\
    \tilde D & \multicolumn{2}{c}{\multirow{2}{*}{\raisebox{\dimexpr\normalbaselineskip+.65\ht\strutbox-1.45\height}[0pt][0pt]
        {\scalebox{2.0}{$L$}}}}\\
    0 & &\\
\end{array} \right]
\\
    B &= \tau\begin{bmatrix}
        0 \\
        I_{g} \\
        0 
    \end{bmatrix},
    & F &= \tau \begin{bmatrix}
        0 & 0\\
        -I_{g} & 0\\
        0 & -I_{n-g} 
    \end{bmatrix},
\end{align*}
where $\tau>0$ is the descritization step size and
\begin{equation*}
    \tilde M = \operatorname{diag}(M_1,\dots,M_{g})^{-1},\quad
    \tilde D = \operatorname{diag}(D_1,\dots, D_{g}).
\end{equation*}
Further, we identify
\begin{align*}
    x(t) &= \begin{bmatrix}
        \omega_\mathcal{G}(t)\\ \theta_\mathcal{G}(t) \\ \theta_{\mathcal{N}\setminus\mathcal{G}}(t)
    \end{bmatrix}\in\mathbb R^{n+g}, & u(t) = p(t)\in\mathbb R^{g}\\
    w(t) &= p^d(t)\in\mathbb R^{n}.&
\end{align*}
We consider the input $w$ as an exogenous variable which cannot be controlled and is predetermined by the actual power demand at the nodes.

\begin{lemma}[Regularity of the descriptor system]~\\
    If the power network described by the susceptance matrix $\bar B$ is connected, then the LDS~\eqref{sysa} is regular, i.e.,\ there exists $\lambda\in\mathbb C$ such that $\det(\lambda E-A)\neq 0$. 
\end{lemma}
\begin{proof}
    For $\lambda\in\mathbb C$ let
    \begin{equation*}
    \tilde L(\lambda) \doteq L-\begin{bmatrix}
        \lambda^2 \tilde M^{-1} +\lambda \tilde D & 0 \\ 0 & 0
    \end{bmatrix}.
    \end{equation*}
    Then with the determinant formula for block matrices
    \begin{align*}
    \label{det}
        &\det\left( \lambda\begin{bmatrix}
    0 & I_g & 0\\
    I_g & 0 & 0\\
    0 & 0 & 0
\end{bmatrix} + \begin{bmatrix}
    I_g & 0 & 0\\
    0 & \tilde M & 0\\
    0 & 0 & I_{n-g}
\end{bmatrix}\left[
\begin{array}{ccc}
    I_g & 0 & 0\\
    \tilde D & \multicolumn{2}{c}{\multirow{2}{*}{\raisebox{\dimexpr\normalbaselineskip+.65\ht\strutbox-1.45\height}[0pt][0pt]
        {\scalebox{2.0}{$L$}}}}\\
    0 & &\\
\end{array} \right]\right)\\
=&\det(\tilde M)\det\left( \lambda\begin{bmatrix}
    0 & I_g & 0\\
    \tilde M^{-1} & 0 & 0\\
    0 & 0 & 0
\end{bmatrix} + \left[
\begin{array}{ccc}
    I_g & 0 & 0\\
    \tilde D & \multicolumn{2}{c}{\multirow{2}{*}{\raisebox{\dimexpr\normalbaselineskip+.65\ht\strutbox-1.45\height}[0pt][0pt]
        {\scalebox{2.0}{$L$}}}}\\
    0 & &\\
\end{array} \right]\right)\\
= &\det(\tilde M)\det(\tilde L(\lambda)).
    \end{align*}
The Laplacian $L$ is a weakly diagonally dominant matrix, that is $\lvert L_{i,i}\rvert \geq \sum_{j\neq i} \lvert L_{i,j}\rvert$ for all $i\in\mathcal N$. Since the power system is connected, for every $i,j\in\mathcal N$ there exists a sequence of nonzero elements $L_{i,i_1}, L_{i_1,i_2},\dots, L_{i_k,j}$. For sufficiently large $\lambda_0$ the diagonally perturbed matrix $\tilde L(\lambda_0)$ has these two properties  as well. Moreover, there exists at least one index $i\in\mathcal N$ such that $\lvert L(\lambda_0)_{i,i}\rvert > \sum_{j\neq i} \lvert L(\lambda_0)_{i,j}\rvert$. This implies the invertability of $\tilde L(\lambda_0)$, cf. \cite{Shivakumar74} and, therefore, $\det(\lambda E-A)\neq 0$ for $\lambda=-\tau\lambda_0+1$.
\end{proof}

\section{Fundamental lemma for Linear Descriptor Systems}

Henceforth, we consider the regular descriptor system \eqref{sysa}, i.e.\ $\det(\lambda E-A)\neq 0$ for some $\lambda\in\mathbb C$. Regularity of~\eqref{sysa} guarantees the existence of invertible matrices $P$,~$S\in\mathbb R^{n\times n}$ which transform the system~\eqref{sysa} into quasi-Weierstraß form (see \cite{BergerIlchmannTrenn12} and \cite[Section~8.2]{Dai89})
\begin{equation}
SEP = \begin{bmatrix}
       I_q & 0 \\ 0 & N
    \end{bmatrix},\quad SAP =\begin{bmatrix}
        A_1 & 0 \\ 0 & I_r
    \end{bmatrix},
\end{equation}
where $A_1\in\mathbb R^{q\times q}$ and $N\in\mathbb R^{(n-q)\times(n-q)}$ is a nilpotent matrix with nilpotency index $s$, i.e.\ $N^{s-1}\neq 0$ and $N^s=0$. Although the quasi-Weierstraß form is not unique, the dimension $q$ and the nilpotency index $s$ do not depend on the particular transformation matrices $P$, $S$ and are, therefore, invariants of the system \eqref{sysa}, cf.\ \cite[Lemma~2.10]{Kunkel00a}. For more details on the properties of LDSs we refer to \cite{Dai89,Stykel02}.

The input-output trajectories of system~\eqref{sys} are collected in the manifest behavior (cf. \cite{PoldermanWillmes98})
\begin{align*}
    &\mathfrak B_\mathrm{m}\doteq\\
    &\left\{ (u,w,y) : \mathbb{N}_0 \rightarrow \mathbb{R}^{g}\times \mathbb R^{n}\times\mathbb{R}^{p} 
    \,\middle|\, \begin{gathered}
        \exists\, x:\mathbb N_0\rightarrow \mathbb R^{n} \text{ s.t.} \\
        x,u,w,y \text{ satisfy~\eqref{sys}}\\
        \text{for all }t\in\mathbb N_0
    \end{gathered}\right\},
\end{align*}
where $\mathfrak B_\mathrm{m}[t,T]\doteq\{b|_{[t,T}\,|\, b\in\mathfrak B_\mathrm{m}\}$ contains the restrictions of the input-output trajectories of \eqref{sys} to the finite time interval $[t:T]$. We say $(u^s,w^s,y^s)\in\mathbb R^{g}\times \mathbb R^{n}\times \mathbb R^{p}$ is an stationary setpoint of the system~\eqref{sys} if $(u,w,y)\in\mathfrak B_\mathrm{m}$ with $u(t)=u^s$, $w(t)=w^s$ and $y(t)=y^s$ for all $t\in\mathbb N_0$.

Next we recall the concepts of R-controllability and R-observability established by \cite{Dai89}, see also \cite{Belov2018,Stykel02}.
\begin{definition}[R-controllability and R-observability]
    The descriptor system \eqref{sys} is called \emph{R-controllable} if
    \begin{subequations}
    \begin{equation}
        \operatorname{rank}\left(\begin{bmatrix}
       \lambda E -A & B & F
\end{bmatrix}\right)=n
    \end{equation}
    holds for all $\lambda\in\mathbb C$.
    System \eqref{sys} is called \emph{R-observable} if
    \begin{equation}
        \operatorname{rank}\left(\begin{bmatrix}
             \lambda E-A\\ C
        \end{bmatrix}\right)=n
    \end{equation}
    \end{subequations}
   holds for all $\lambda\in \mathbb C$.\End
\end{definition}

\noindent We recall the notion of persistency of excitation. 
\begin{definition}[Persistency of excitation]
    A function $u:[0:T-1]\rightarrow \mathbb R^{m}$ is said to be \emph{persistently exciting} of order~$L$ if the Hankel matrix $H_{L} (\mathbf{u}_{[0,T-1]})$ has rank $mL$, i.e., full row rank.\End
\end{definition}

The following result for LDSs from \cite[Lemma 5]{Schmitz2022} lays the foundation of the present paper.
\begin{lemma}[Fundamental lemma for LDSs]
\label{mainth}
    Suppose that the system~\eqref{sysa} is R-controllable and regular. Let $(\bar u, \bar w, \bar y)\in \mathfrak B_{\mathrm m}[0,T-1]$ be such that $\begin{bmatrix}\bar u^\top, \bar w^\top\end{bmatrix}{}^\top$ is persistently exciting of order $L+q+s-1$ and $T,L \in \mathbb{N}$ satisfy $(g+n+1)(L+q+s)-1\leq T$. Then $(u,w,y)\in\mathfrak B_\mathrm{m}[0,L-1]$ if and only if there is $\alpha\in\mathbb R^{(g+n+p)L\times (T-s-L+2)}$ such that
    \begin{equation}
    \label{dynamic}
        \begin{bmatrix}
            H_L(\bar{\mathbf{u}}_{[0,T-s]})\\
            H_L(\bar{\mathbf{w}}_{[0,T-s]})\\
            H_L(\bar{\mathbf{y}}_{[0,T-s]}) 
        \end{bmatrix}\alpha = \begin{bmatrix}
           \mathbf{u}_{[0,L-1]}\\ \mathbf{w}_{[0,L-1]} \\ \mathbf{y}_{[0,L-1]}
        \end{bmatrix}.
    \end{equation}
    \End
\end{lemma}
Lemma~\ref{mainth} gives rise to a non-parametric description of system~\eqref{sys}, where no explicit knowledge of the matrices $E,A,B,F,C$ is needed. Every element of the manifest behavior $\mathfrak B_\mathrm{m}[0,L-1]$, that is every input-output trajectory of system~\eqref{sys}, corresponds to some vector $\alpha$ and vice versa. In particular, an identification step for the system matrices $E,A,B,F,C$ is not necessary to reconstruct the trajectories of system~\eqref{sys}.

\section{Data-driven predictive control}

Next we apply Lemma~\ref{mainth} to an Optimal Control Problem~(OCP) and propose a data-driven MPC scheme. Suppose that the system~\eqref{sysa} is regular and \eqref{sys} is R-controllable and R-observable.

Given an observed trajectory $(u,w,y)\in\mathfrak B_\mathrm{m}[t-q-s+1,t-1]$ up to time $t-1$ and a (predicted) constant future power demand $p^d\in\mathbb R^{n}$, we consider the OCP
\begin{subequations}
\label{ocp_org}
\begin{equation}
    \label{ocp_org1}
    \operatorname*{minimize}_{(\hat u,\hat w,\hat y)}~ 
    \sum_{k=0}^{L-1}  \lVert\hat y(t+k)-y^s\rVert_{Q}^2 + \lVert\hat u(t+k)-u^s\rVert_{R}^2
\end{equation}
subject to $(\hat u,\hat w,\hat y)\in\mathfrak B_\mathrm{m}[t-q-s+1, t+L-1]$ and 
\begin{align}
    \label{ocp_org3}
    \begin{bmatrix}
           \hat{\mathbf{u}}_{[t-q-s+1,t-1]} \\
           \hat{\mathbf{w}}_{[t-q-s+1,t-1]}\\
           \hat{\mathbf{y}}_{[t-q-s+1,t-1]}
        \end{bmatrix} &= \begin{bmatrix}
           \mathbf{u}_{[t-q-s+1,t-1]}\\ \mathbf{w}_{[t-q-s+1,t-1]} \\ \mathbf{y}_{[t-q-s+1,t-1]}
        \end{bmatrix}, \\
    \label{ocp_org4}
    \begin{bmatrix}
         \hat{\mathbf{u}}_{[t+L-q-s+1, t+L-1]}\\
         \hat{\mathbf{y}}_{[t+L-q-s+1,  t+L-1]}
    \end{bmatrix}
    &=\begin{bmatrix}
         u^s\\\vdots\\u^s\\y^s\\\vdots\\y^s
    \end{bmatrix},\\
    \label{ocp_org5}
    \hat{\mathbf{w}}_{[t,t+L-1]} &= \begin{bmatrix}
         p^d\\\vdots\\p^d
    \end{bmatrix},\\
    \label{ocp_org6}
    \hat{\mathbf u}_{[t,t+L-1]} &\geq 0,
\end{align}
\end{subequations}
where $(u^s,p^d,y^s)$ is an stationary setpoint. The matrices $Q\in \mathbb R^{p\times p}$ and $R\in\mathbb R^{m\times m}$ in the stage cost are assumed to be symmetric and positive-definite. The stage cost penalizes deviations from $y^s$, i.e.\ the desired output, and the control effort. The consistency constraint~\eqref{ocp_org3} guarantees the latent state of the predicted and true trajectory are aligned up to time $t-1$, cf. \cite{Schmitz2022}. The constraint \eqref{ocp_org5} ensures that the power demand  $p^d$ is met. The nonnegativity established by the inequality constraint~\eqref{ocp_org6} serves to match the physical interpretation of the variable $\hat u$ as mechanical power provided to the generators.

For the prediction horizon $L$ we assume that  $L\geq \tilde L + q+s-2$, where $\tilde L = 2s+q$. This together with the terminal constraint~\eqref{ocp_org4} implies initial as well as recursive feasibility of OCP~\eqref{ocp_org} and that the stationary setpoint $(u^s, p^d, y^s)$ is asymptotically stable with respect to the optimal control in closed loop, at least when the constraints \eqref{ocp_org5} and \eqref{ocp_org6} are neglected, cf.\ \cite[Proposition~10]{Schmitz2022}. Imposing the latter constraints, the initial feasibility has to be ensured explicitly, while in this case recursive feasibility and stability follow automatically.

Lemma~\ref{mainth} implies that all trajectories contained in the manifest behavior~$\mathfrak B_\mathrm{m}[t-q-s+1,t-1]$ can be parameterised by a Hankel matrix. Hence, given an input-output trajectory $(\bar u, \bar w, \bar y)\in\mathfrak B_\mathrm{m}[0,T-1]$ with persistently exciting input $\begin{bmatrix}\bar u^\top, \bar w^\top\end{bmatrix}{}^\top$ of order $L+2(q+s-1)$, OCP~\eqref{ocp_org} is equivalent \citep{Schmitz2022} to 
\begin{subequations}\label{ocp}
\begin{equation}
    \label{ocp1}
    \operatorname*{minimize}_{(\hat u,\hat w,\hat y,\alpha(t)), }\sum_{k=0}^{L-1} \lVert\hat y(t+k)\rVert_{Q}^2 + \lVert\hat u(t+k)\rVert_{R}^2
\end{equation}
with $(\hat u,\hat w,\hat y):[t-q-s+1: t+L-1]\rightarrow \mathbb R^{g }\times \mathbb R^{n}\times \mathbb R^{p}$ and $\alpha(t)\in\mathbb R^{T-L-2s-q+3}$ subject to
\begin{align}
    \label{ocp2}
    \begin{bmatrix}
           \hat{\mathbf{u}}_{[t-q-s+1,t+L-1]} \\
           \hat{\mathbf{w}}_{[t-q-s+1,t+L-1]}\\
           \hat{\mathbf{y}}_{[t-q-s+1,t+L-1]}
        \end{bmatrix} &= \begin{bmatrix}
            H_{L+q+s-1}(\bar{\mathbf{u}}_{[0,T-s]}) \\
            H_{L+q+s-1}(\bar{\mathbf{w}}_{[0,T-s]})\\
            H_{L+q+s-1}(\bar{\mathbf{y}}_{[0,T-s]})
        \end{bmatrix}\alpha(t),\\
    \label{ocp3}
    \begin{bmatrix}
           \hat{\mathbf{u}}_{[t-q-s+1,t-1]} \\
           \hat{\mathbf{w}}_{[t-q-s+1,t-1]}\\
           \hat{\mathbf{y}}_{[t-q-s+1,t-1]}
        \end{bmatrix} &= \begin{bmatrix}
           \mathbf{u}_{[t-q-s+1,t-1]}\\ \mathbf{w}_{[t-q-s+1,t-1]} \\ \mathbf{y}_{[t-q-s+1,t-1]}
        \end{bmatrix}, \\
    \label{ocp4}
    \begin{bmatrix}
         \hat{\mathbf{u}}_{[t+L-q-s+1, t+L-1]}\\
         \hat{\mathbf{y}}_{[t+L-q-s+1,  t+L-1]}
    \end{bmatrix}
    &=\begin{bmatrix}
         u^s\\\vdots\\u^s\\y^s\\\vdots\\y^s
    \end{bmatrix},\\
    \label{ocp5}
    \hat{\mathbf{w}}_{[t,t+L-1]} &= \begin{bmatrix}
         p^d\\\vdots\\p^d
    \end{bmatrix},\\
    \label{ocp6}
    \hat{\mathbf u}_{[t,t+L-1]} &\geq 0,.
\end{align}
\end{subequations}

In our data-driven MPC strategy, OCP~\eqref{ocp} is solved at each time step~$t$ and, for the solution $(u^\star, w^\star, y^\star, \alpha^\star(t))$, the value~$u^\star(t)$ is applied as new input $u(t)$ to the system~\eqref{sys}.
The MPC scheme based on the OCP~\eqref{ocp} we propose for the descriptor system~\eqref{sys} is summarized in Algorithm~\ref{algo}.

\begin{algorithm}
\caption{\textbf{: Data-enabled predictive control}}
\textbf{Input:} prediction horizon~$L$, (pers.\ exciting) input/output data $(\bar u,\bar w,\bar y)$
\label{algo}
\begin{algorithmic}[1]
\State Set $t=0$
\State Measure $(u,w,y)\in\mathfrak B_\mathrm{m}[t-q-s+1,t-1]$ and predict the future power demand $p^d$
\State Compute $(u^\star, w^\star, y^\star, \alpha^\star(t))$ to  \eqref{ocp}
\State Apply $u(t)=u^\star(t)$
\State $t\gets t+1$ \quad and\quad goto Step\,$2$
\end{algorithmic}
\end{algorithm}
\section{Numerical Example}

\begin{figure}[htb]
	\centering
	\includegraphics[width=0.6\linewidth]{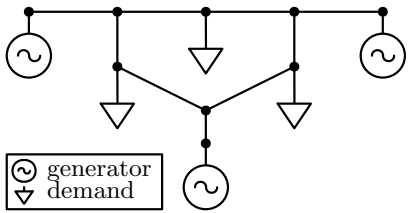}
	\caption{Nine-bus power system with three generators and three consumers.}
	\label{fig:9bus}
\end{figure}

We consider a nine-bus system \citep{Schulz1977}, with parameters from \cite{Cole2011a}.
The system comprises three generators ($g=\lvert \mathcal G\rvert = 3$) and three consumers, see Fig.~\ref{fig:9bus}. 

The corresponding linearized and temporal discretized descriptor system \eqref{sysa} is regular ($q=6$, $s=1$) as well as R-controllable. For the output we assume that only data of the phase angles at the generator nodes is accessible, i.e. 
\begin{equation}
    C = \begin{bmatrix} 0_{3\times 3} & I_3 &  0_{3\times 6}\end{bmatrix},\quad y=\theta_{\mathcal G}=\begin{bmatrix}\theta_1 & \theta_2 & \theta_3\end{bmatrix}^\top.
\end{equation} 
With this choice of $C$ the system~\eqref{sys} is R-observable. 

We compare the performance of the data-driven control approach described by Alogorithm~\ref{algo} and a classical droop controller realized via a feedback loop 
\begin{equation}
     p = \tilde p - K\omega_{\mathcal G},
\end{equation}
where $\tilde p\in\mathbb R^{3}$ is some constant offset.

\begin{figure}[t]
    \centering
    \includegraphics[scale=0.64]{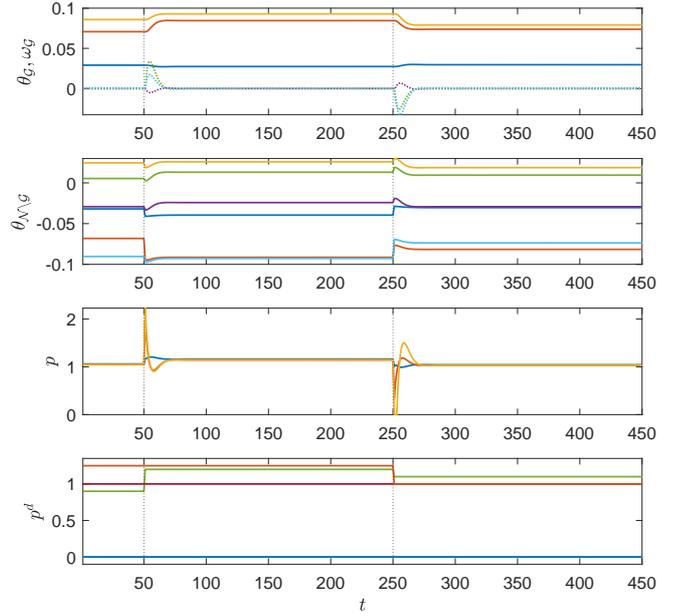}
    \caption{A trajectory emerging from the data-driven predictive control scheme presented in Algorithm~\ref{algo}. The power demand changes at times $t=50$ and $t=250$. }
    \label{fig:ddpc}
\end{figure}

Given a certain power demand $p^d = \begin{bmatrix}
    p_1^d & \dots & p_9^d
\end{bmatrix}$, the aim is to steer the system into a corresponding stationary setpoint $(u^s, p^d, y^s)$ of the system \eqref{sys}.
We apply the predictive control Algorithm~\ref{algo} with prediction horizon $L=20$. As a persistently exciting input signal of order $L + 2(q+s-1) = 32$ we choose a function $\bar u:[0: 399]\rightarrow \mathbb R^{12}$ with values drawn independently from a uniform distribution over the interval $[-1,1]$. The matrices in the stage cost function are chosen as $Q=10\cdot I_3$ and $R=I_3$.
Fig.~\ref{fig:ddpc} illustrates a trajectory generated by Algorithm~\ref{algo}, where two changes in the demand of power occur. %
\begin{figure}[t]
	\centering
	\includegraphics[scale=0.64]{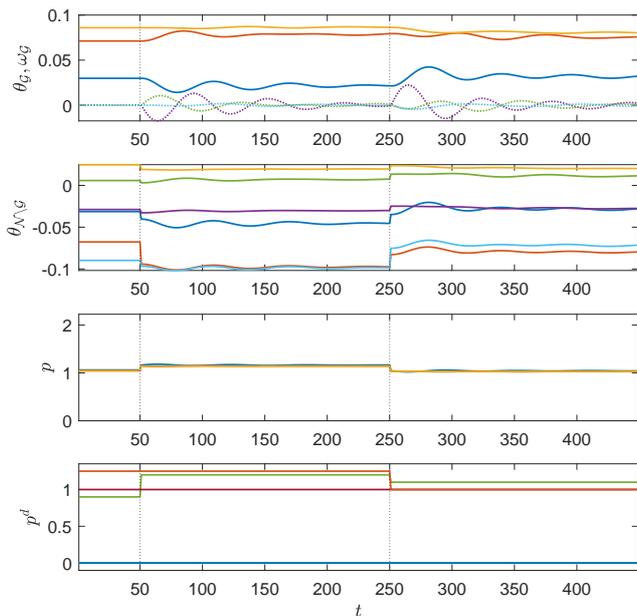}
	\caption{The figure depicts a trajectory corresponding to a droop controller, where changes in the power demand occur at time $t=50$ and $t=250$.}
	\label{fig:droop}
\end{figure}
Fig.~\ref{fig:droop} shows the closed-loop trajectory corresponding to a droop controller. 
Observe that compared to the classical droop controller, the data-driven MPC controller yields a faster frequency stabilization with smaller oscillations. 
This comes at the cost of a large alteration of the mechanical power on the generators down to their lower bounds, which might not be desirable for synchronous generators due to the risk of mechanical stress.
However, in case of mainly inverter-based future power grids, such a control action can be safely applied  leading to a better overall control performance.

\section{Conclusion}

In the present work we have proposed an approach for data-driven MPC for frequency stabilization with stability guarantees.
Our simulation results show a promising performance of the used controller compared to conventional droop control. 

Our current model is linear and relies on assumptions such as constant bus voltages and small voltage angle differences, which can be restrictive in practice. 
Hence, future work aims at relaxing these assumptions. Moreover, the consideration of stochastic uncertainties is subject to future work.

%

\bibliography{bibAlex, refs}             
                                                   







\end{document}